# Nonlinear energy harvesting

F. Cottone[1], L. Gammaitoni[1,2*], H. Vocca[2]

[1] *N.i.P.S. Laboratory, Dipartimento di Fisica, Università degli Studi di Perugia - 06100 Perugia, Italy.*

[2] *INFN Sezione di Perugia, Via A. Pascoli, 1 – 06100 Perugia, Italy*

[*]*To whom correspondence should be addressed. Email: luca.gammaitoni@pg.infn.it*

**Ambient energy harvesting has been in recent years the recurring object of a number of research efforts aimed at providing an autonomous solution to the powering of small-scale electronic mobile devices[1-3]. Among the different solutions, vibration energy harvesting[4-7] has played a major role due to the almost universal presence of mechanical vibrations: from ground shaking to human movements, from ambient sound to thermal noise. Standard approaches are mainly based on resonant linear oscillators that are acted on by ambient vibrations. Here we propose a new method based on the exploitation of the dynamical features of stochastic nonlinear oscillators. Such a method is shown to outperform standard linear oscillators and to overcome some of the most severe limitations of present approaches, like narrow bandwidth, need for continuous frequency tuning and low efficiency. We demonstrate the superior performances of this method by applying it to piezoelectric energy harvesting from ambient vibration. Experimental results from a toy-model oscillator are described in terms of nonlinear stochastic dynamics. We prove that the method proposed here is quite general in principle and could be applied to a wide class of nonlinear oscillators and different energy conversion principles. There are also potentials for realizing micro/nano-scale power generators[8].**

Ambient vibrations come in a vast variety of forms from sources as diverse as wind induced movements, seismic noise and cars motion. Present working solutions for vibration-to-electricity conversion are based on oscillating mechanical elements that convert kinetic energy via capacitive, inductive or piezoelectric methods[9-11]. Mechanical oscillators are usually designed to be resonantly tuned to the ambient dominant frequency. However, in the vast majority of cases the ambient vibrations have their energy distributed over a wide spectrum of frequencies, with significant predominance of low frequency components and frequency tuning is not always possible due to geometrical/dynamical constraints[11-14].

To overcome these difficulties we propose a different approach based on the exploitation of the properties of non-resonant oscillators. Specifically we demonstrate that a bistable oscillator, under proper operating conditions[15], can provide better performances compared to a linear oscillator in terms of the energy extracted from a generic wide spectrum vibration. In order to do so, we realized a toy-model oscillator made by a piezoelectric inverted pendulum (See Supplementary Fig. 1) where on top of the pendulum mass it has been added a small magnet (tip magnet). The effect of ground vibration force is reproduced by applying a properly designed magnetic excitation on two small magnets attached near the base of the pendulum. Under the action of the excitation the pendulum oscillates, alternatively bending the piezoelectric beam and thus generating a measurable voltage signal. The dynamics of the inverted pendulum tip can be controlled with the introduction of an external magnet conveniently placed at a certain distance $\Delta$ and with polarities opposed to those of the tip magnet. The external magnet introduces a force dependent from the distance $\Delta$ between the two magnets that opposes the elastic restoring force of the bended beam and, as a result, the inverted pendulum dynamics can show two different types of dynamics. When the external magnet is far away, the inverted pendulum behaves like a linear oscillator whose dynamics is resonant with a resonance frequency determined by the system parameters. This situation accounts well for the usual operating condition of traditional piezoelectric vibration-to-electric energy converters[6]. On the other hand, when $\Delta$ is small enough two new equilibrium positions appear. The random vibration makes the pendulum swing in a more complex way with small oscillations around each of the two equilibrium positions and large excursions from one to the other.

In order to quantify the energy produced by the piezoelectric oscillator we computed the power dissipated in a purely resistive load, by measuring the voltage drop $V$ over a resistive load $RL$[7], under the influence of a random vibration with Gaussian distribution (with zero mean and standard deviation $\sigma$) and exponential autocorrelation function (with correlation time $\tau$). In Fig. 1 (upper) we show the average electrical power $<V^2>/RL$ as a function of $\Delta$ for three different values of the noise standard deviation $\sigma$. In all the cases the power increases rapidly from the linear case (large $\Delta$) up to a maximum value and then decreases when the magnets become closer and closer. A qualitatively similar behaviour is observed (Fig. 1, lower) if we plot the pendulum rms position $x_{rms}$, as a function of $\Delta$. In order to quantitatively account for the experiments we developed a dynamical description of the inverted pendulum based on the following equation of motion: $m\ddot{x} = -U'(x) - \gamma\dot{x} - K_v V(t) + \sigma\xi(t)$. The first term on the right hand side accounts for the conservative force, where $U(x)$ is the potential energy of the pendulum[16] shown in Fig.2: $U(x) = Kx^2 + (ax^2 + b\Delta^2)^{-3/2} + c\Delta^2$ with $a,b,c$ representing constants related to the physical parameters of the pendulum[17,18] (See Supplementary Fig. 1). The second term $-\gamma\dot{x}$, accounts for the energy dissipation due to the bending and $-K_v V(t)$ accounts for energy transferred to the electric load R with coupling equation $\dot{V} = K_c \dot{x} - V/RC$. C and $K_c$ are respectively the capacitance and the coupling constant of the piezoelectric sample. Finally, $\sigma\xi(t)$ accounts for the vibration force that drives the pendulum. $\xi(t)$ represents a stochastic process with the same statistical properties of the magnetic excitation. In Fig. 1 we plot with continuous line the computed power $<V^2>/R$ (upper) and the rms value of $x(t)$ (lower) as obtained by the numerical solution of the equation of motion. All the parameters have been measured from the experimental apparatus and introduced into the equation. As it is apparent the agreement between the experimental data and the model is rather good.

In fig. 1 we can easily identify three different regimes: 1) Large $\Delta$, (i.e.. $\Delta >> \Delta_c = \alpha/\sqrt{b}$ with $\alpha = (3a/2K)^{1/5}$). The dynamics is characterized by quasi-linear oscillations around the single minimum located at zero displacement, in correspondence with the vertical position of the pendulum. This condition accounts for the usual performances of a linear piezoelectric generator. 2) Small $\Delta$ ($\Delta << \Delta_c$), the potential energy is bistable with a very pronounced barrier between the two wells. In this condition and for a given amount of noise, the pendulm swing is almost exclusively confined within one well and the dynamics is characterized once again by quasi-linear oscillations around the minimum of the confining well. 3) In between there is a range of distances $\Delta$ where the $x_{rms}$ (and the $V_{rms}$ as well) reaches a maximum value. In this condition the pendulum dynamics is highly nonlinear and the swing reaches its largest amplitude with noise assisted jumps between the two wells. As it is well evident in fig. 1, the maximum values of the output power exceed by a factor that ranges between 4 and 6 the value obtainable when the magnet is far away. This indicates a potential gain for power harvesting between 400% and 600% compared to the standard linear oscillators, depending on the noise intensity and on the other physical features of the pendulum. Two others important features are apparent: a) the maximum position shifts toward larger $\Delta$ when the noise intensity increases. b) In the low $\Delta$ regime, the rms reaches a plateau that is smaller than the plateau reached by the rms in the large $\Delta$ regime.

Both features can be explained as follows. When $\Delta >> \Delta_c$ the potential $U(x)$ shows a single minimum with librational frequency given by $\omega_0^2 = U''(0)/m = \left(2k - 3ab^{-5/2}/\Delta^5\right)/m$. The $x_{rms}$ value here can be estimated in the linear oscillator approximation[19] as proportional to $\sigma/\omega_0$. On decreasing $\Delta$, $\omega_0$ decreases and thus the $x_{rms}$ value increases. When $\Delta = \Delta_c$ the potential develops two distinct minima located at $x_\pm = \pm\sqrt{(\alpha^2 - b\Delta^2)1/a}$. The pendulum swings now between the two minima and the rms increases proportional to $x_\pm$. With decreasing $\Delta$ the potential barrier height $\Delta U$ grows proportional to $\Delta^{-3}$ and becomes so large that the jump probability becomes negligible. The pendulum swing is thus permanently confined within one well. Such a trapping condition happens at smaller values of $\Delta$ (i.e. larger barrier) for larger noise. This explains the observed shift of the maximum position toward smaller $\Delta$ as observed in a). Inside one well the dynamics is almost linear with small oscillations around the potential local minima ($x_+$ or $x_-$) and $x_{rms} \propto \sigma/\omega_\pm$ with $\omega_\pm^2 = U''(x_\pm)/m$. Being $\omega_\pm > \omega_0$ it follows $rms_{\Delta \to 0} < rms_{\Delta \to +\infty}$, as observed in b).

The increase in the rms value observed for the inverted pendulum is not a peculiar feature of this specific system or potential. Instead it appears to be a quite general feature of bistable dynamical systems in the presence of noise. To support this statement we focussed our attention on the dynamics of the so-called Duffing oscillator[20], extensively studied in the presence of noise both in the classical[21] and in the quantum domain[22]. The potential $U_q(x) = -a/2\, x^2 + b/4\, x^4$ is bistable when $a > 0$ with $x_\pm = \pm\sqrt{a/b}$ and $\Delta U_q = a^2/4b$. The equation of motion: $m\ddot{x} = -U_q'(x) - \gamma_q \dot{x} + \sigma_q \xi(t)$ is in analogy with the pendulum case. The role of $\Delta$, is played here by the parameter $a$. The behaviour of $x_{rms}$ is qualitatively similar to the one shown by the pendulum (See Supplementary Fig. 2), i.e. three distinct regimes can be identified: 1) $a << 0$. The potential is monostable and the dynamics is characterized by quasi-linear oscillations around the minimum located at $x = 0$. 2) $a >> 0$. The potential is bistable with a very pronounced barrier between the two wells. The dynamics is mainly trapped inside one minimum. 3) In between there is a range of values of *a* where $x_{rms}$ reaches a maximum and the dynamics is characterized by frequent jumps between the two wells. In Fig. 3 we present the $x_{rms}$ as a function of *a* and the noise variance $\sigma^2$. The contour plot view in the figure inset shows the evolution of the maximum $x_{rms}$. The solid line is a theoretical prediction obtained with the following argument. The rms evolution in regime 3, can be roughly modelled as governed by to two main contributions: i) the raising, mainly due to the growth of the separation between the two minima $x_{\pm}$; ii) the drop, mainly due to the decrease in the jump probability measured by the crossing probability, proportional[23] to $\exp(-\Delta U_q/\sigma^2\tau)$, caused by the increase of the potential barrier height $\Delta U_q$. For the sake of identifying the dependence of the maximum position we computed the root of the equation $d(x_{rms})/da = d/da\left(x_+ \cdot \exp(-\Delta U_q/\sigma^2\tau)\right) = 0$ obtaining that $x_{rms}$ reaches a maximum when $a = a_{max} = \sqrt{b\sigma^2\tau}$ for a given noise intensity.

Finally we would like to stress the fact that the results obtained can led to a significant increase in energy harvesting performances also in the small scale domain, where the dynamical features discussed here can be applied without changes also to micro[24,25] and nanomechanical resonators[26]. Energy generators based on nanoscale linear oscillators have been proposed[8] and noise driven dynamics are considered as a promising option[27,28].

**Acknowledgements**

Useful discussions with P. Amico, I. Neri and F. Marchesoni are gratefully acknowledged.

**Correspondence and requests for materials should be addressed to L.G. (***luca.gammaitoni@pg.infn.it***)**.

Figure 1   Figure 1 | **Piezoelectric oscillator mean electric power (upper panel) and position $x_{rms}$ (lower panel) as a function of Δ for three different values of the noise standard deviation σ.** The symbols correspond to experimental values measured from the apparatus described in Supplementary Fig.1. The piezoelectric material used in the pendulum is Lead Zirconate Titanate. The continuous curves have been obtained from the numerical solution of the stochastic differential equation described in the text. Both in the experiment and in the numerical solution, the stochastic force has the same statistical properties: Gaussian distribution, with zero mean and standard deviation σ, exponentially auto-correlated with correlation time τ = 0.1 s. Every data point is obtained from averaging the rms values of ten time series sampled at a frequency of 1 KHz for 200 s. The rms is computed after zero averaging the time series. The expected relative error in the numerical solution is within 10%.

Figure 2   Figure 2 | **Inverted pendulum potential function U(x).** Five different plots of the potential function $U(x) = Kx^2 + (ax^2 + b\Delta^2)^{-3/2} + c\Delta^2$ corresponding to five different values of the parameter Δ, representing the distance between the external magnet and the tip magnet (See Supplementary Fig.1) are shown (vertical scale in J). The constants $K,a,b,c$ are related to the pendulum parameters through: $K = K_{eff}/2$ with $K_{eff}$ effective elastic constant; $a = \left(\mu_0 M^2/2\pi d\right)^{-2/3} d^2$ with $\mu_0$ the permeability constant, $M = 0.051$ Am$^2$, the effective magnetic moment and $d = 2.97$ a geometrical parameter related to the distance between the measurement point and the pendulum length; $b = a/d^2$ and $c = K/d^2$. On decreasing Δ the potential changes from monostable to bistable. The distance between the two minima located in $x_\pm$ increases when Δ decreases, while the barrier height increases.

Figure 3   Figure 3 | **Root mean squared (rms) values of the Duffing oscillator position $x_{rms}$ as a function of $a$ and the noise variance $\sigma^2$.** The values plotted here have been obtained from the numerical solution of the stochastic differential equation described in the text. The rms is computed after zero averaging the $x(t)$. In the inset: the contour plot view of $x_{rms}$. Superimposed on the plot there is a continuous line representing the theoretical prediction for the maximum: i.e. $x_{rms}$ has a maximum when $a = \sqrt{b\sigma^2\tau}$. The expected relative error in the numerical solution is within 10%.

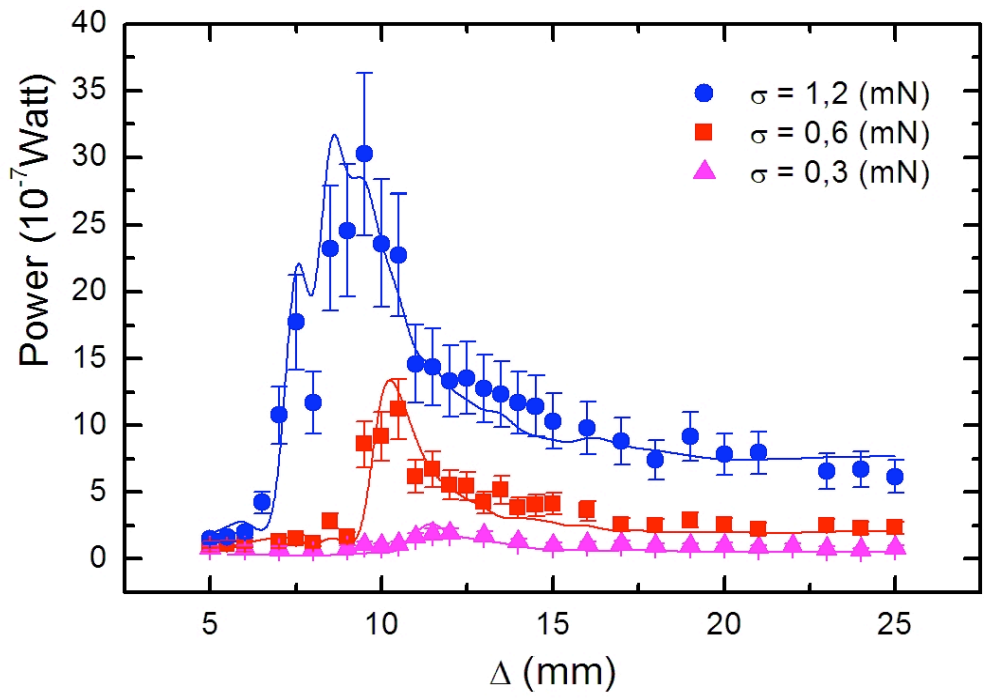

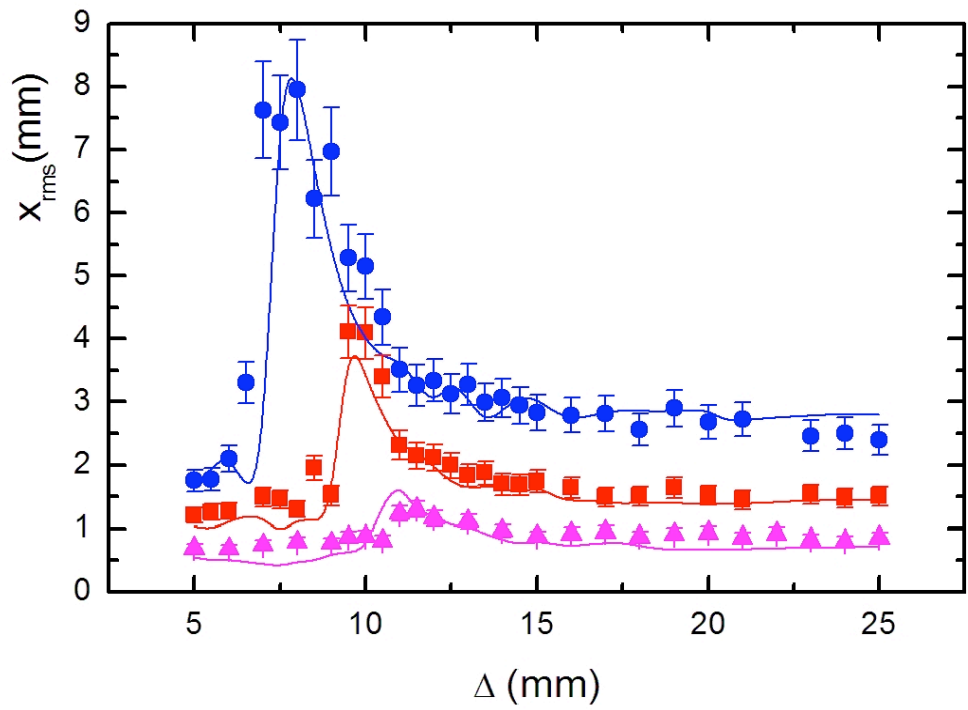

Fig. 1

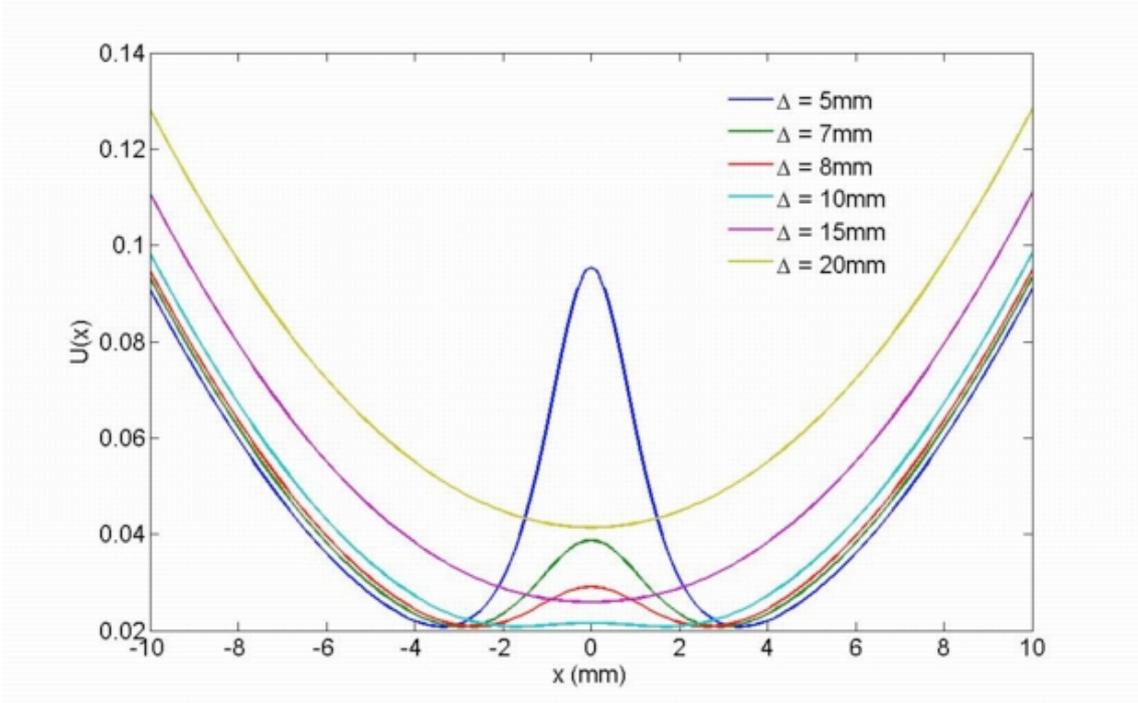

Fig.2

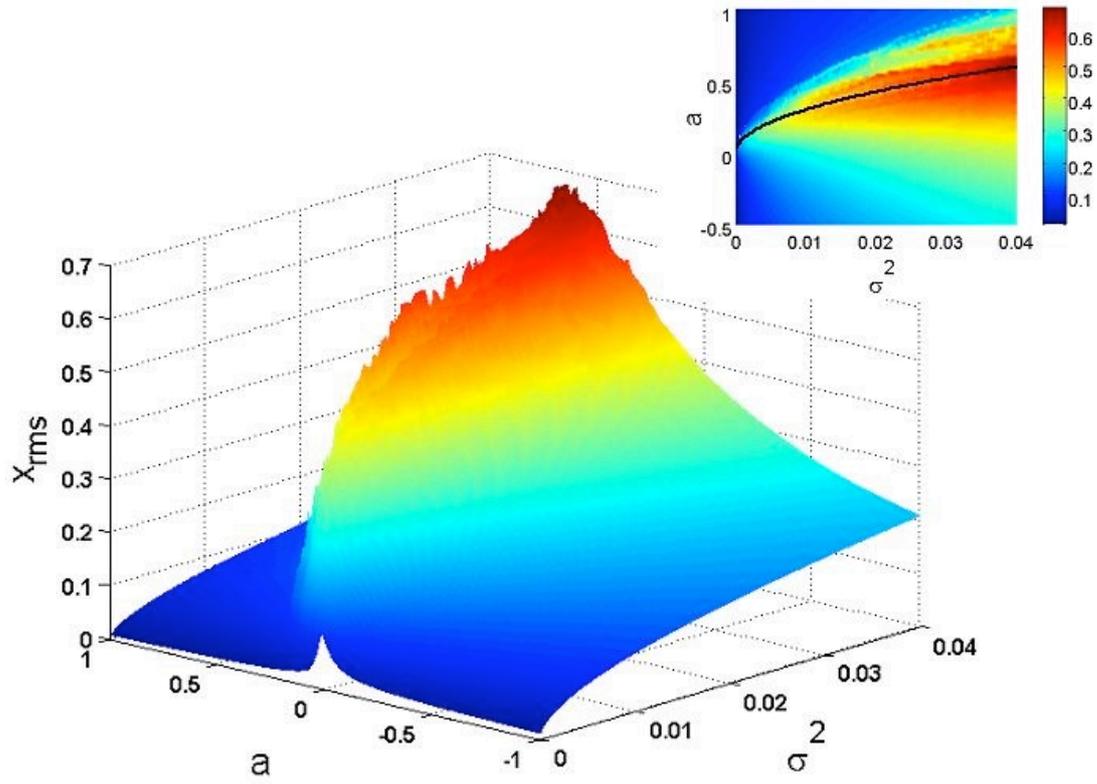

Fig.3

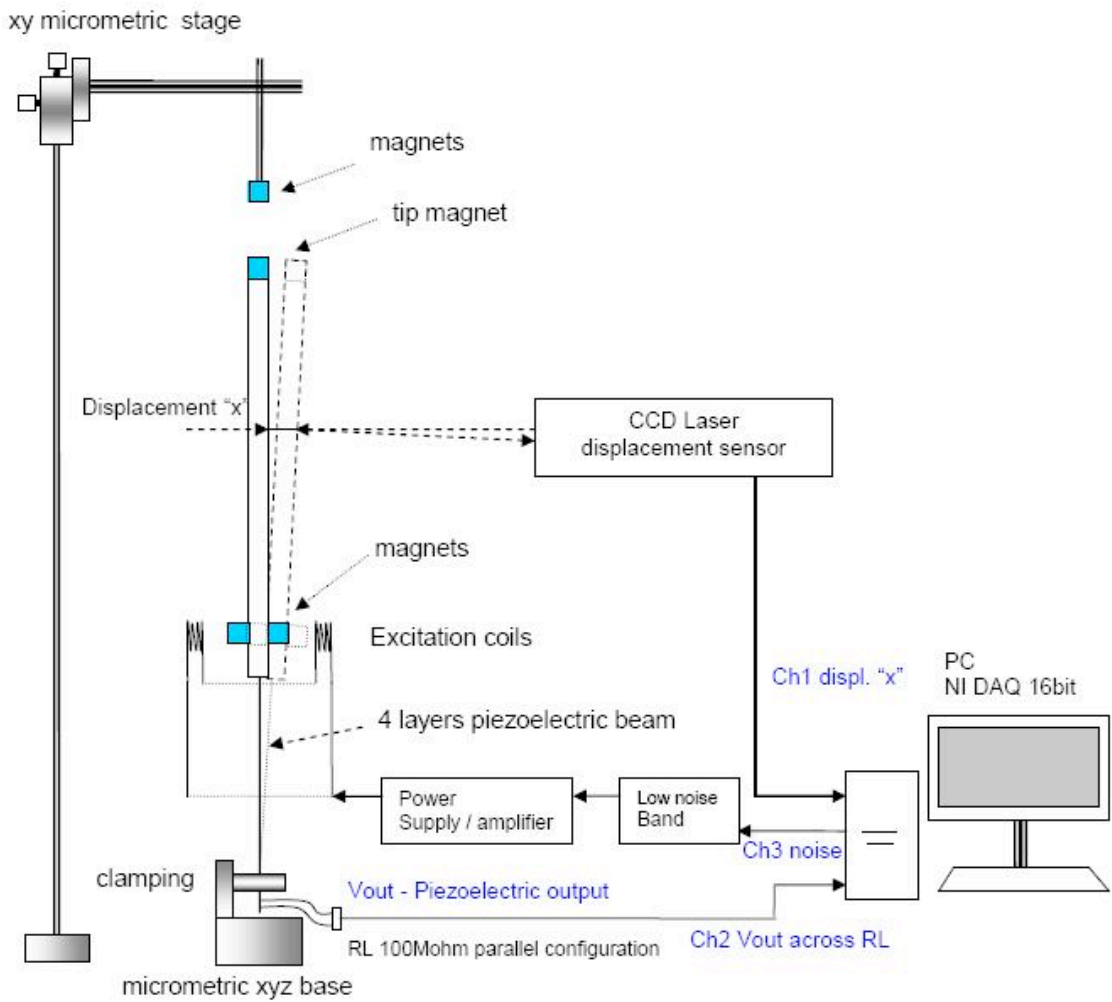

**Supplementary Information. Figure 1. Experimental apparatus.**
The figure presents a schematic of the apparatus employed in the experiment. The inverted pendulum is realized with a four-layer piezoelectric beam (mod. T434-A4-302
4-Layer Bender by Piezo system inc.) made by Lead Zirconate Titanate (PSI-5A4E) 60 mm of free length, clamped at one end. The piezoelectric beam has a width of 5 mm and a thickness of 0.86 mm. The pendulum base position can be adjusted via a micrometric xyz displacement system. The pendulum mass is a steel cylinder 140.0 mm long and with diameter of 4.0 mm, with attached three magnets (each magnetic dipole moment = 0.051 $Am^2$ ). The inverted pendulum resonance frequency (in the linear regime) is 6,67 Hz. The tip magnet is faced by a similar magnet with inverse polarities placed at a distance D and held in place by a massive structure. The distance D can be adjusted via a micrometric displacement control system. The displacement x is measured via an optical read-out with a CCD-Laser displacement sensor with sensitivity of the order of 0.05 mm. The signal from the displacement sensor is sampled by a digital signal processing board (DSPB) controlled by a personal computer with sampling frequency 1KHz. The voltage signal from the piezo is measured through a load resistor RL placed in parallel with piezoelectric output voltage terminals and sampled by the DSPB. The digitalized signals are stored in the computer memory for post-processing elaboration. The DSPB is employed also to drive a current generator that produces, through a couple of coils, the magnetic excitation that mimic the ground vibration. The signal generated by the DSPB is filtered and conditioned in order to reproduce the desired statistical properties. The actual magnitude of the standard deviation of the vibrational force applied in the three cases is s= 3 $10^{-4}$ N, 6 $10^{-4}$ N, 12 $10^{-4}$ N. The load resistance is R=RL = 100 MOhm and the piezo capacitance C=112 nF. The effective mass is m=0.0155 kg. The damping constant is $\gamma$=0.016 Hz.

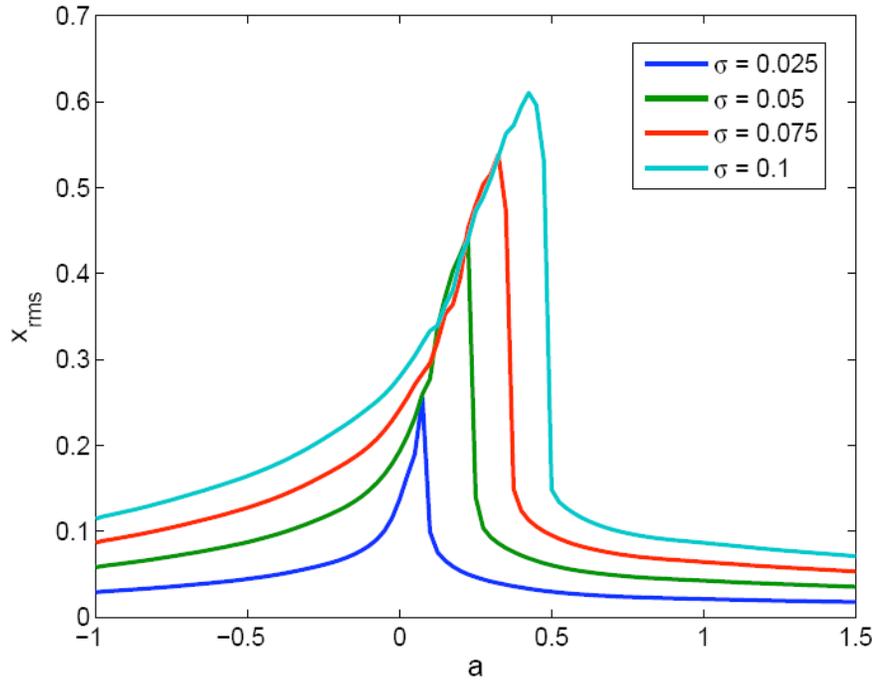

**Supplementary Information. Figure 2 | Root mean squared (rms) values of Duffing oscillator as a function of *a* for four different values of the noise standard deviation $\sigma_q$.** The continuous curves have been obtained from the numerical solution of the stochastic differential equation $m\ddot{x} = -U_q'(x) - \gamma_q \dot{x} + \sigma_q \xi(t)$ with the potential $U_q(x) = -a/2\, x^2 + b/4\, x^4$. The stochastic force $\sigma_q \xi(t)$ has the same statistical properties as for the pendulum potential case: Gaussian distribution, with zero mean and standard deviation $\sigma_q$, exponentially auto-correlated with correlation time $\tau = 0.1$ s. The rms is computed after zero averaging the relevant time series. The expected relative error in the numerical solution is within 10%.